# *Contextual guidance:*
# *An integrated theory for astrocytes' function in brain circuits and behavior*

PERSPECTIVE ARTICLE

**Ciaran Murphy-Royal [1], ShiNung Ching [2], Thomas Papouin [3†]**

[1] Centre de Recherche du Centre Hospitalier de l'Université de Montréal (CRCHUM) & Département de Neurosciences, Université de Montréal, Canada

[2] Electrical and Systems Engineering, Washington University in St. Louis, MO, USA

[3] Department of Neuroscience, Washington University School of Medicine, St. Louis, MO, USA

† For correspondence: thomas.papouin@wustl.edu

**Abstract –** The participation of astrocytes in brain computation was formally hypothesized in 1992, coinciding with the discovery that these glial cells display a complex form of $Ca^{2+}$ excitability. This fostered conceptual advances centered on the notion of reciprocal interactions between neurons and astrocytes, which permitted a critical leap forward in uncovering many roles of astrocytes in brain circuits, and signaled the rise of a major new force in neuroscience: that of glial biology. In the past decade, a multitude of 'unconventional' and disparate functions of astrocytes have been documented that are not predicted by these canonical models and that are challenging to piece together into a holistic and parsimonious picture. This highlights a disconnect between the rapidly evolving field of astrocyte biology and the conceptual frameworks guiding it, and emphasizes the need for a careful reconsideration of how we theorize the functional position of astrocytes in brain circuitry. Here, we propose a unifying, highly transferable, data-driven, and computationally-relevant conceptual framework for astrocyte biology, which we coin contextual guidance. It describes astrocytes as contextual gates that decode multiple environmental factors to shape neural circuitry in an adaptive, state-dependent fashion. This paradigm is organically inclusive of all fundamental features of astrocytes, many of which have remained unaccounted for in previous theories. We find that this new concept provides an intuitive and powerful theoretical space to improve our understanding of brain function and computational models thereof across scales because it depicts astrocytes as a hub for circumstantial inputs into relevant specialized circuits that permits adaptive behaviors at the network and organism level.

## 1- *In search of a holistic framework for astrocyte biology*

Following the discovery that astrocytes, a type of glia, possess a form of intracellular $Ca^{2+}$ excitability strikingly slower than that of neurons [1], Stephen Smith hypothesized that "astrocyte networks might mediate slow modulations of neuronal function," citing arousal, selective attention, mood, and learning as examples [2]. These early predictions, proven true over the subsequent decades, embraced the slowness of astrocytes' excitability as a distinctive feature. Noting that "brain function, in all its complexity and glory, requires many different kinds of computation – some discrete, fast and specific; others slow and diffuse", Smith de facto positioned astrocytes as a slow computing unit of the brain [2].

Three decades later, the question "*what do astrocytes do in the brain?*" still has no simple answer, only a list of disparate functions fulfilled by these cells in the CNS, from ion homeostasis to synapse regulation, energy supply, and blood flow control [3]. This highlights a lack of general and communicable agreement on why astrocytes fundamentally exist in the brain and the need for a unifying and broadly applicable theory that portrays astrocytes' participation to CNS functions across scales and species, in transferable and useful terms. To this end, several conceptual hypotheses were formulated in the past that enabled considerable strides in elucidating astrocytes contribution to synaptic function, hemodynamics and circuit activity, and still guide the field so far as laying its modern foundations [3]. Two popular examples are the lactate shuttle hypothesis [4], in which astrocytes are depicted as an energetic intermediate between neuronal activity and blood flow, and the tripartite synapse hypothesis, in which astrocytes are thought to monitor and influence activity at individual synapses in their domain [5,6]. While broadly accepted, these traditional perspectives are limited to a single scale and do not articulate, together or separately, an overarching vision of astrocyte function. This is evidenced by a recent wave of new discoveries that are incompletely captured by these standard views and appear to challenge the conceptual horizon of the field (**Table 1**), as noted by others [7]. To address the rising disconnect between the rapidly evolving field of astrocyte biology and the conceptual frameworks guiding its enquiry, there needs to be a profound reconsideration of how we theorize the functional position of astrocytes in the brain.

We reason that, to be valid, a unifying framework for astrocyte biology must satisfy the following criteria: 1) Consistency: it must reflect past and recent observations; 2) Parsimony: it must provide simple explanations to a variety of phenomena with few theoretical parts; 3) Usefulness: it should shed new light on existing challenges, controversies or inconsistencies, thereby opening new perspectives; 4) Robustness: it must continue to satisfy the first three criteria as new discoveries are made.

An ideal astrocyte theory would thus capture all fundamental features of astrocytes, preserved across species: their morphological complexity [8], tiling [9,10], gap-junction coupling [11], interaction with blood vessels [12], and molecular heterogeneity [13,14]. It would, at the functional level, logically articulate astrocytes' slowness [15,16], their responsiveness to a variety of signals and conditions not predicted by existing models (**Table 1**), their insensitivity to the blockade of neuronal activity documented by many (but not all) labs [17,18], the roles they play at different sub-cellular locations, and the variety of factors they release [3,19-21]. To be parsimonious, it would only rely on few reductionist principles, the most undisputed of which are astrocytes' position as sentinels of the interstitial space, owing to their high surface-to-volume ratio, intricate integration in the neuropil [8], and versatile molecular makeup [3], as well as their ability to dynamically orchestrate the physical and chemical properties of the extracellular milieu (or "active milieu" [7]) including the activity and connectivity of the neuropil [3,20,21]. Finally, one way that an astrocyte theory would be useful is in combining the contribution of astrocytes with that of high-speed processing by neurons in producing cognition and behavior [21].

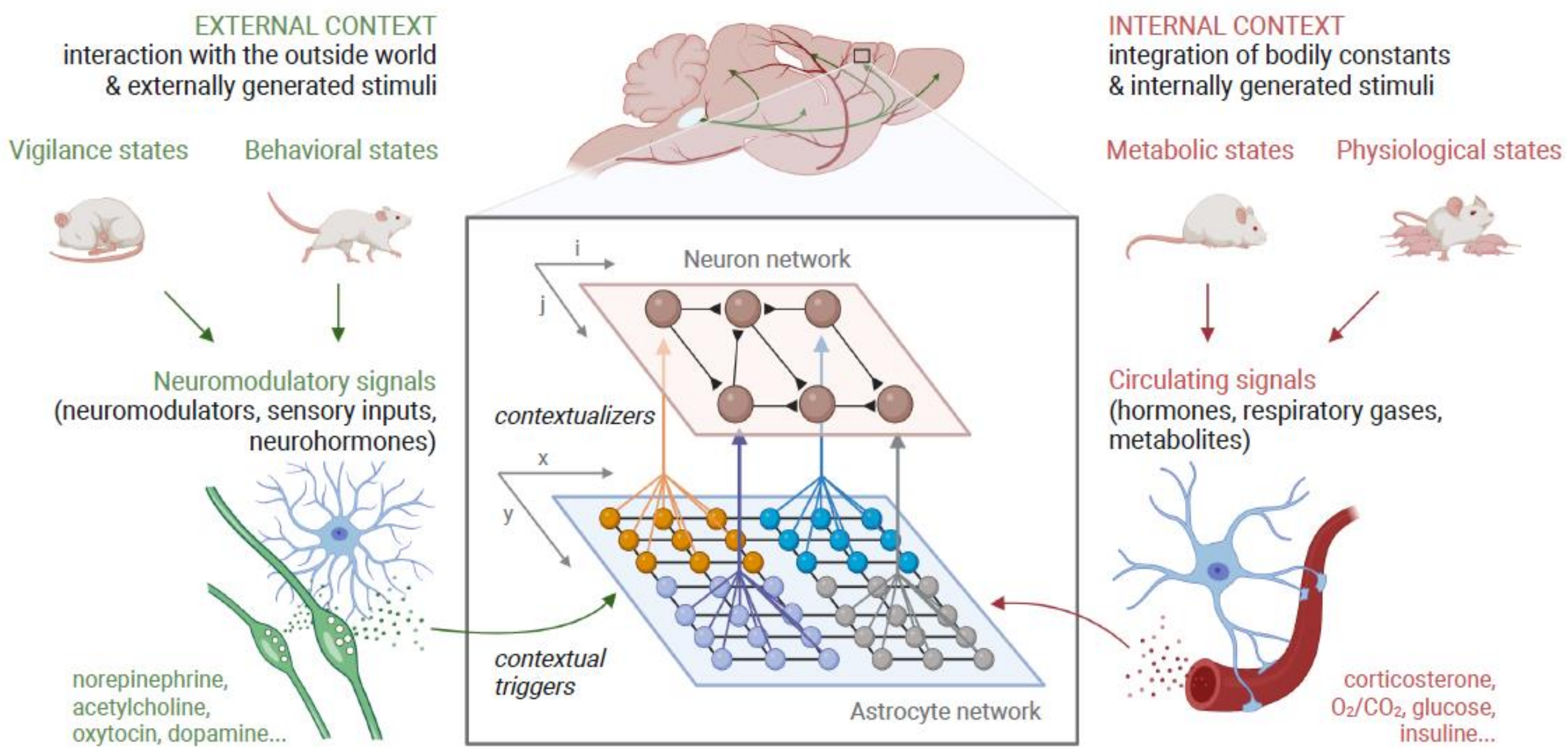

***Figure 1:*** **Astrocyte networks are a separate layer of information processing cued by contextual inputs to guide neural circuits**. Neuronal networks and astrocytic networks operate as separate layers, the former operating rapid point-to-point transfer and integration of information, and the latter transducing slow modulatory information over large spatial scales. In this system, astrocytes are the primary receiver of slow contextual signals that carry information about the status of the animal with respect to the outside world. Due to their privileged and systematic interaction with blood vessels, astrocytes are also the first responders to circulating signals that carry information about the animal bodily status. In response to these various stimuli, astrocyte networks signal onto neuronal elements to alter circuit connectivity and/or activity. The net result is a context-specific adaptive reconfiguration of the neural network topology and functional output. Also illustrated is the gap-junction coupling of individual astrocytes into functional networks with coordinated outputs (see main text).

## 2- *Contextual guidance: astrocytes configure neural circuits to ongoing contexts*

With these guidelines, we here introduce *contextual guidance*, an intuitive and universal theory for astrocytes as state-dependent orchestrators of neural circuitry (**Fig. 1**). In brief, it states that 1) astrocytes are tuned to circuit-relevant organismal contexts by sensing state-dependent cues; 2) these cues mobilize astrocytic outputs that reconfigure the activity and/or topology of the underlying circuit; and 3) these modifications produce context-specific adaptation of the neural network.

In this three-step framework, astrocytes are first (**Fig. 2A**, step 1) mobilized by signals that i) diffuse through the interstitial space (e.g., [22]), ii) bear significance to the function of the local circuit (e.g., [23]), and iii) encode a change in animal's state (e.g., [24]). We term these signals *contextual triggers* because they convey information on the organismal status and mobilize astrocyte signaling (**Box 1**). We give examples of such signals from the literature in **Table 1**. By definition, contextual triggers are circuit-specific and will vary greatly in nature, prevalence, and origin, across the CNS, yielding a seemingly heterogeneous repertoire of signals to which astrocytes respond, some shared across the brain (e.g., norepinephrine, NE), others unique to select nuclei (e.g., $pCO_2$ [25]). This fully captures the non-uniformity of astrocytes' responsiveness to many types of signaling molecules in the brain, satisfying criterion 1, Consistency.

In response to contextual triggers, (step 2) astrocytes mobilize effectors that are consequential to the local circuitry and can take many forms, including synapse-bound transmitters [26], lactate supply [27], glutamate uptake [28], extracellular $Ca^{2+}$ and $K^{+}$ regulation [29,30], synapse elimination [31], or retraction/extension of peri-synaptic astrocytic processes (PAPs, [32,33]), to name a few. We coin these astrocytic outputs *contextualizers* because they alter neural properties in response to a context (**Box 1**, and step 3 below). Contextualizers are thus defined based on their circuit-modifying competence, rather than their nature, mode of mobilization, or mechanism of action, reflecting the diversity of astrocyte outputs that abound in the literature and fulfilling criterion 1, Consistency.

Finally, (step 3) the net effect of contextualizers is to shape the activity (excitability, firing mode, synchrony) or functional topology (synaptic wiring, fidelity, potency, efficacy, plasticity) of the neuropil, as abundantly evident from the literature of the past 30 years (criterion 1, Consistency). We call these circuit modifications *contextual tuning* because, importantly, we postulate that they produce functional adaptation to the context (**Box 1**). Instances of the full contextual guidance flow and its adaptive nature already exist in the literature and are summarized in **Table 2**. For example, during high-vigilance states (*context*), cholinergic signaling (*contextual trigger*) drives astrocytic D-serine release (*contextualizer*) to augment NMDAR readiness at CA3-CA1 synapses (*contextual tuning*), which enhances the potential for hippocampal learning during periods of spatial exploration (*adaptation*) [26].

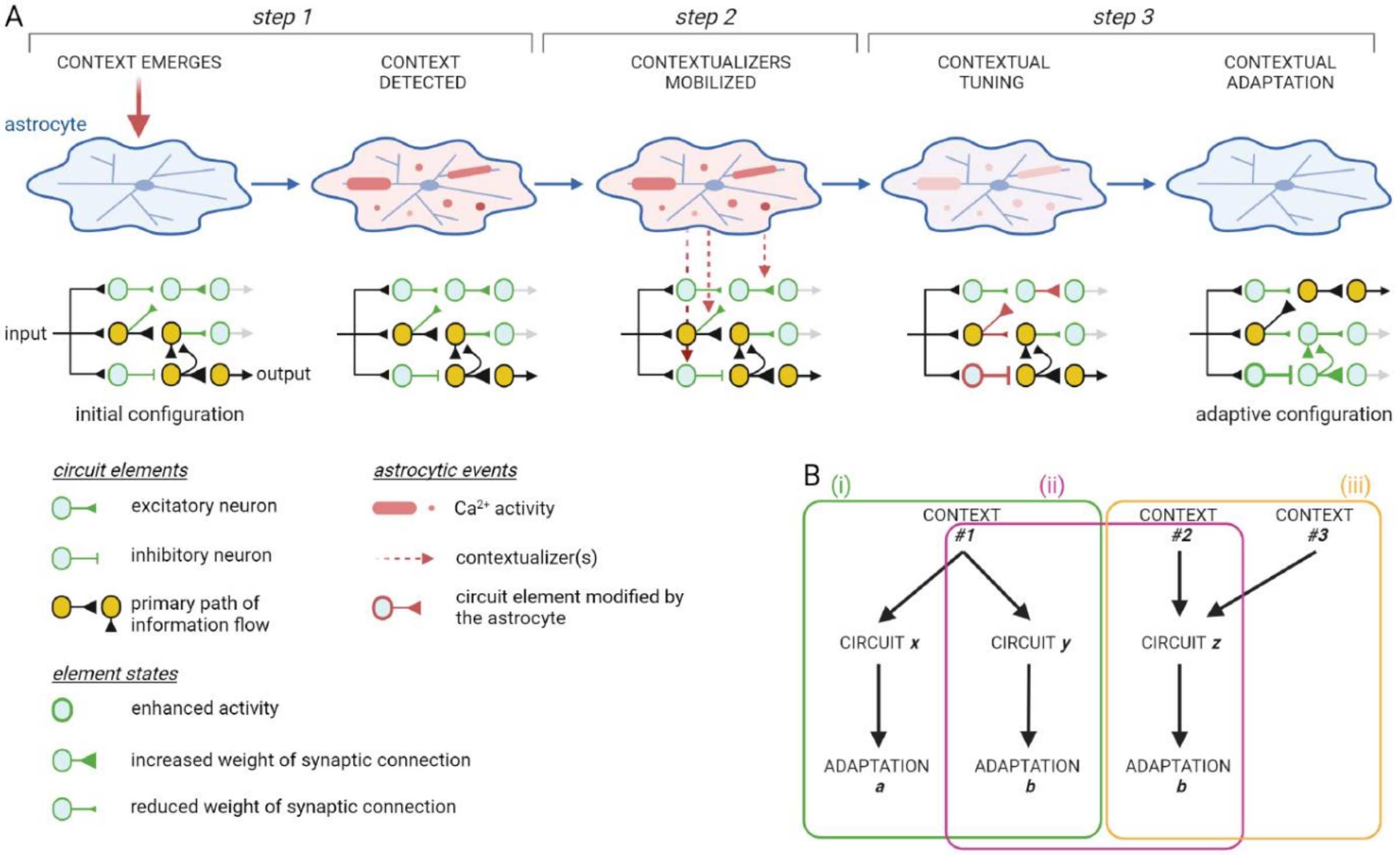

***Figure 2:*** **Astrocytes implement circuit reconfigurations in response to changing context**. **A**, Stepwise illustration of the principles of contextual guidance, showing an individual astrocyte and embedded neuronal circuit over time (with typical micro-circuit motifs). The astrocyte is recruited by the emergence of a context (shown as an elevation in $Ca^{2+}$ activity). This leads to the mobilization of circuit-modifying contextualizer(s), which act on select elements of the circuit (here synaptic weight as in [97], and interneuron firing as in [64]) to modify the functional connectivity and activity of the overall network. Under this new configuration, network computation is reshaped, illustrated as a change in the path of least resistance through the circuit (shown in black) and main output. **B**, Schematic illustrating different scenarios in which (i) a context, #1, yields circuit adaptations *a* and *b* in circuits *x* and *y* respectively, (ii) two contexts, #1 and #2, yield the same adaptation *b* in separate circuits *y* and *z*, and (iii) adaptation *b* is achieved in circuit *z* in response to either context, #2 or #3, (via the same or different contextualizers).

**Box 1** | **Definitions**

**Context**: ensemble of conditions that relate to an internal or external organismal status, such as a physiological, metabolic, biochemical, behavioral, or vigilance state

**Contextual trigger**: a signal that i) diffuses through the interstitial space, ii) bears significance to the function of the local circuit, iii) encodes a change in internal or external state and, iv) has the ability to mobilize astrocyte signaling

**Contextualizer**: astrocytic output, mobilized in response to a contextual trigger, that is consequential to the underlying neural circuitry

**Contextual tuning**: effect of a contextualizer on the activity and/or functional topology of a neural circuit

**Contextual adaptation**: adaptative value of the circuit reconfiguration produced under contextual tuning, with regard to the context that triggered it

**Contextualizing rules**: association of specific instances of the elemental parts described above into a unique "context→contextual trigger→contextualizer→contextual tuning and adaptation" sequence

The notion of *context* is here defined in alignment with mounting evidence that astrocytes are attuned to brain states (**Table 1**). Hence, “context” refers to the conditions that arise with a physiological [32], metabolic [28], biochemical [23], behavioral [34], or vigilance state [35]. An essential commonality of “contexts”, therefore, is that they relate to an internal or environmental status of the animal. With this view, astrocytes become a hub for circumstantial inputs onto relevant specialized circuits (**Fig. 1**). This is remarkably consistent with the features of early astrocyte homologues, CEPsh glia in C. elegans nerve-ring, which extend processes that associate with sensory neuron-receptive endings of the animal's nose, forming post-synaptic-like structures sensitive to environmental cues [36].

## 3- *Unifying astrocytes' roles and features in one simple paradigm*

In the following sections, we illustrate the usefulness of contextual guidance in articulating disparate observations and extracting new notions, in fulfillment of criteria 2, Parsimony and 3, Usefulness.

### *a. Untangling astrocytes inconsistent input-output rules across the CNS*

In the contextual guidance theory, there are no constraints *a priori* on the possible combinations of contextual triggers and contextualizers. On the contrary, it allows a contextual trigger to mobilize distinct contextualizers across various circuits, and vice versa, which provides a consensual ground for seemingly inconsistent results in the literature. For instance, (i) while NE signaling elicits the release of astrocytic D-serine in the mouse spinal cord [37], it has been linked to the release of ATP in the cortex [38]. Conversely, (ii) astrocytic D-serine secretion is gated by NE in the spinal cord [37], by oxytocin (OT) in the amygdala [22], and by acetylcholine (ACh) in the cortex and hippocampus [26,39]. Finally, (iii) D-serine availability in the CA1 is also governed by other pathways, such as endocannabinoid signaling [40]. In the contextual guidance theory, these disparities simply reflect the fact that *contextualizing rules* (what context yields what adaptation, **Box 1**) differ from circuit to circuit, such that a context can yield different adaptations in circuits that serve different functions (example (i) above), distinct contexts can lead to the same adaptation in independent circuits (ii), and within a circuit, multiple contexts may trigger the same adaptation (iii) (**Fig. 2B**).

Contextual guidance also allows a contextualizer to have varied effects across regions, and possibly ages or disease states, as these effects ultimately depend on downstream neural targets. For instance, the secretion of the $Ca^{2+}$-chelator S100β by astrocytes causes a drop in the extracellular $[Ca^{2+}]$. In the masticatory central pattern generator (CPG), this elicits the bursting of neurons by de-inactivating a conductance called $I_{Nap}$ [29] while, in the primary visual cortex layer 5, this triggers the spiking of pyramidal cells by activating Nav1.6 sodium channels [41].

These considerations illustrate the usefulness of contextual guidance in disentangling inconsistent reports of astrocytes' input-output rules, that is, the signals that recruit astrocytes across the brain (see also **3-e.**), the signaling pathways engaged in response to them, and their downstream effects on neural circuits.

### *b. From astrocytes' heterogeneity to circuit-based functional specialization*

Astrocytes' heterogeneity is now well-documented at the transcriptomic and proteomic level, and with respect to morphology, synaptic coverage, domain volume, electrophysiological properties, $Ca^{2+}$ activity and functions, across and within brain-regions [13,14]. For instance, striatal and hippocampal astrocytes share a backbone of commonly expressed genes related to core astrocytic functions, but differentially express 10% of genes/proteins, including $K^+$ channels, $Ca^{2+}$ regulators, and $Ca^{2+}$-binding effectors. This yields distinct molecular signatures, accompanied by a number of functionally relevant differences [17]. Recent reports extended the notion of heterogeneity to finer levels, such as across cortical layers [42-44], echoing intra-striatal nuances [17]. Concomitant differences in morphological complexity and synaptic ensheathment, here too, led to the hypothesis that astrocytes engage in subtly different functions across layers. Hence, this type of work has not only revealed the complex landscape of astrocytes' attributes across the brain, down to the circuit level, it also provides strong precedent that the brain comprises myriad neural-circuit specialized astrocytes rather than a homogeneous population [45]. The present theory embraces this idea in that it requires such molecular versatility to generate astrocyte phenotypes dedicated to circuit-relevant contextual tuning (**Table 2**). In fact, we hypothesize that astrocytes' heterogeneity reflects the repertoire of contextualizing rules at play across brain circuits. Since distinct rules can be obtained with minor variations in commonly expressed genes (e.g., those governing $Ca^{2+}$ excitability) and a handful of uniquely expressed genes (e.g., the thyroid hormone binding protein μ-crystallin in striatal astrocytes [17]), this could explain the combinatorial diversity of molecular identities across the brain, along a gradual continuum, rather than discrete astrocyte subclasses.

### *c. In contextual guidance, astrocytes' tiling is a compartmentalization mechanism*

The contextual guidance theory can shed light on other features of astrocytes that remain overlooked, including why astrocytes tile the entire brain in individual non-overlapping domains. This hallmark, preserved from nematodes to primates and humans [9,10,21], is unaccounted for in most models because it is still largely unexplained at the functional level. Contextual guidance postulates that, in response to context-bearing signals, astrocytes alter the neuropil within their sphere of influence. A corollary is that a mechanism must exist to guarantee that, within an astrocyte domain, only one set of contextualizing rules applies. With such premises, the tiling of astrocytes naturally imposes itself as a compartmentalization mechanism that ensures the spatial separation of individual functional units (**Fig. 3**). In effect, the absence of overlap between neighboring astrocytes, along with their ubiquitousness, permits the reliable implementation of unambivalent contextualizing rules across the entire CNS.

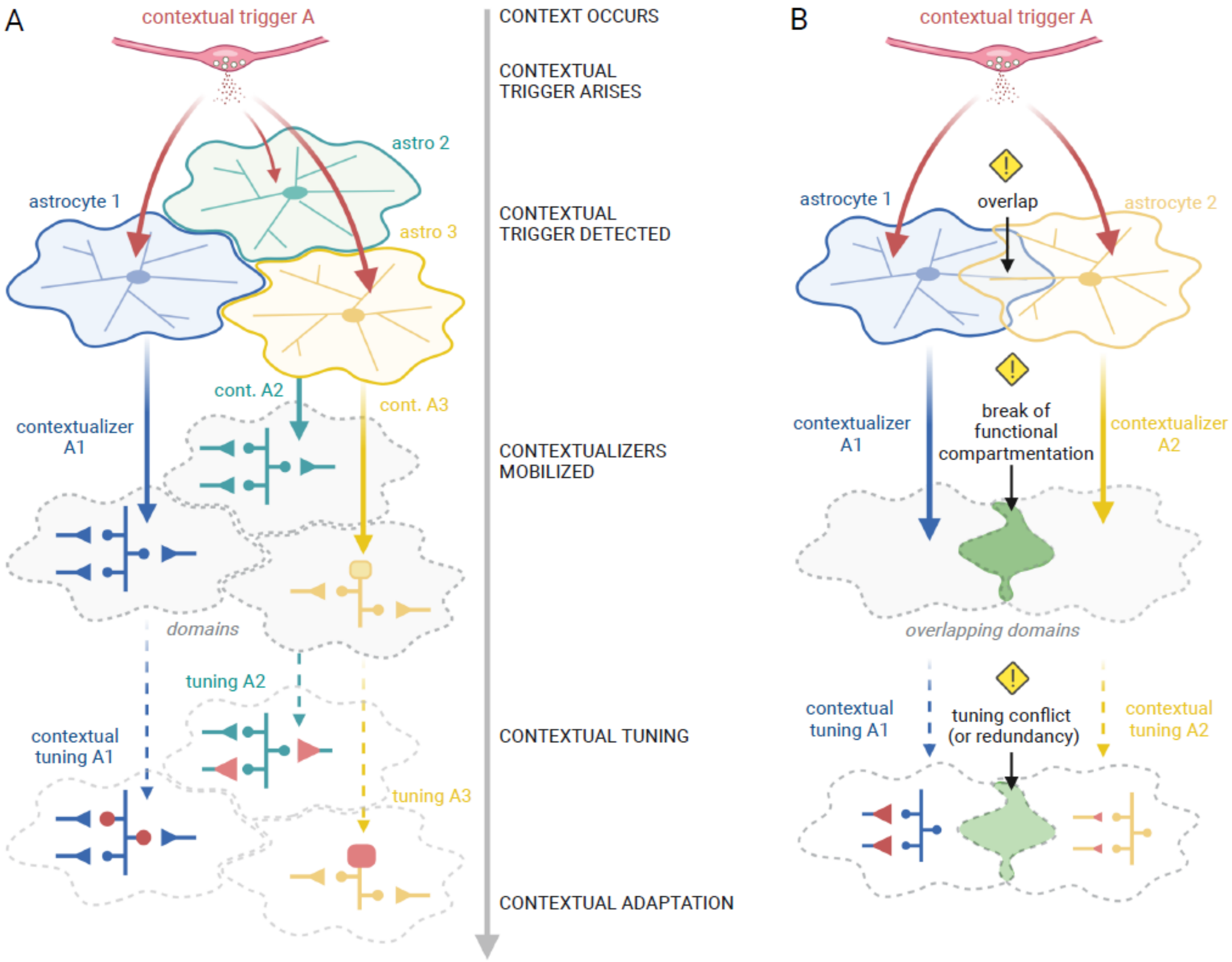

***Figure 3:*** **Astrocytes' tiling is required for the spatial separation of functional compartments**. **A,** The tiling of astrocytes ensures that a single set of contextualizing rules applies to any element of the neuropil at any given time. In effect, this is a compartmentalization mechanism necessary for the implementation of circuit-specific rules with high fidelity and specificity. This is illustrated by three astrocytes that, upon detection of context A, each deploy a unique contextualizer in their domain, to achieve different contextual tunings and circuit-specific adaptations (respectively increase in synaptic potency, efficacy or neuronal excitability [30-32]). **B,** Hypothetical scenario wherein two neighboring astrocyte territories overlap. This leads to a compartmentalization breach, wherein neuronal elements in the overlapping zone are subjected to two sets of potentially conflicting contextualizing rules (here shown as different contextualizers in response to context A). This may result in functional redundancies if the rules are synergetic, or conflicts if the rules are antagonistic (shown here: contextualizers 1A and 2A augment and reduce synaptic efficacy, respectively).

### *d. Gap-junction coupling extends the spatial boundaries of contextualizing rules*

Just as the contextual guidance theory provides functional significance to the anatomical tiling of astrocytes, it can be applied to understand the functional importance of astrocyte networks. It has been known for decades that astrocytes share ions and small molecules (<1kDa) such as water, glucose, lactate, ATP, $Ca^{2+}$ or $K^+$, through an extensive network of gap-junctions formed by connexin hemichannels [11]. Remarkably, the boundaries of such networks vary from brain region to brain region and can mirror the basic boundaries of the underlying cytoarchitecture [46,47]. But the biological logic behind the existence of these networks and their precise anatomy remains elusive. In the present paradigm, we can hypothesize that a network of gap-junction coupled astrocytes reflects a constellation of astrocytes sharing the same contextualizing rules, effectively connecting functional units that operate identically. The core function of gap-junction coupling of astrocytes into networks might thus be three-fold:

(1) Enlarge the receptive field of individual cells by allowing the concerted detection of contexts across coalescent astrocytes, and provide resilience to the system (if one astrocyte fails, the network copes). A corollary is that a subset of astrocytes interspersed in the network may act as primary sensors of a given context and entrain the rest of the network in the contextualizing response. Remarkably, evidence for such phenomena already exists, such as in the amygdala, where the responsiveness of the astrocyte network to OT is entrained by a subset of morphologically distinct, OTR-positive astrocytes [22].

(2) Extend the jurisdiction of a set of contextualizing rules and coordinate a unified response over large neural circuits. This is well-illustrated by the role of astrocytes in orchestrating functional state changes such as sleep [35,48-52], which involves the concerted action of many astrocytes to alter $[K^+]_{ex}$, $[Mg^{2+}]_{ex}$, $[Ca^{2+}]_{ex}$ and pH, lactate supply, interstitial space volume and glymphatic circulation [53-55].

(3) Separate ensembles of astrocyte governed by independent, potentially distinct rules and modalities of circuit adaptation. This implies that astrocyte networks act to segment the brain into volumes of neural tissue that adapt differently to organismal contexts (**Fig. 3**).

### *e. What signals do astrocytes really respond to? Finding the tune in a cacophony*

The heterogeneous range of molecules now known to influence astrocytes' activity has been challenging to reconcile with traditional perspectives [4-6] because many of these stimuli do not arise from the local network (**Table 1** & **Fig. 4**). The contextual guidance theory, on the contrary, is modeled after these very observations, (listed in **Tables 1 & 2**), which, by virtue of generalization, yield the notion that astrocytes are tuned to circuit-relevant organismal conditions at large.

In doing so, the contextual guidance theory largely excludes the possibility that astrocytes are entrained by the unitary activity of all hundreds of thousands of individual synapses in their territory, because information about the state of the animal is often not relayed by local synapses but by cocktails of circulating signals or peripheral afferents. This starkly departs from the tripartite synapse concept [6,56] in which baseline synaptic activity is considered the primary driver of astrocyte signaling, such that astrocytes perform a “re-encoding of fast synaptic information that feeds back onto neurons for retransmission of that information” [2]. Intriguingly, the computational and evolutionary advantages of such feedback have remained elusive and the timing of these interactions has been problematic owing to the slowness of astrocyte signaling compared to synapses [15,16 and refs within], putting into question the fundamental relevance of synapse-triggered astrocytic outputs in such framework (**Fig. 4**). Moreover, in many cases, blocking neuronal activity has little to no effect on astrocyte $Ca^{2+}$ dynamics [17,18]. Perhaps this indicates that salient and coordinated synaptic inputs, rather than unitary synaptic transmission, may inform astrocytes of high local activity and energy demand, and help tune the astrocytic response to the network state, as hinted by others [57]. Evidence indeed suggests that astrocytes’ recruitment by synapses often necessitates strong or synchronized activity [58]. Such conditions favor transmitter spillover and diffusive retrograde signaling (e.g. endocannabinoids) both of which fulfill the contextual trigger criteria and are known to elicit astrocyte signaling [40,58,59]. There is also a growing appreciation for astrocytes’ sensitivity to inhibitory synapses [60,61], and evidence that discrete populations of inhibitory contacts that play salient roles in reshaping the activity of neuronal ensembles and orchestrating network outputs [62] can entrain astrocytes to influence circuit activity [63,64]. If true, this defies the decades-long idea that astrocytes maintain synaptic homeostasis via a feedback loop, in favor of one where they act as a sensor/effector element in a feedforward mechanism where the adjusted variable is the adequacy of the local circuit configuration and the stimulus is the ongoing context (**Fig.4**).

Contextual guidance can also shine light on signals that may not have functional relevance to astrocytes. Thalamic visual inputs, in V1 of the mouse cortex, may be such an example. Indeed, owing to their raw nature in this early stage of visual processing and continuous incidence during wakefulness, they do not fulfill the third contextual trigger criterion (“encode a change in internal or external state”, **Box 1**). Information about the context in which these inputs occur, however, such as the motion of the subject with respect to the visual scene, might be relevant to how they are processed ("are the horizontal bars moving or am I moving through a landscape of static horizontal bars?") [65]. Consistent with these considerations, Paukert et al. demonstrated that V1 astrocytes are largely insensitive to visual stimuli but respond strongly to bouts of locomotion and supra-linearly to the pairing of both [34].

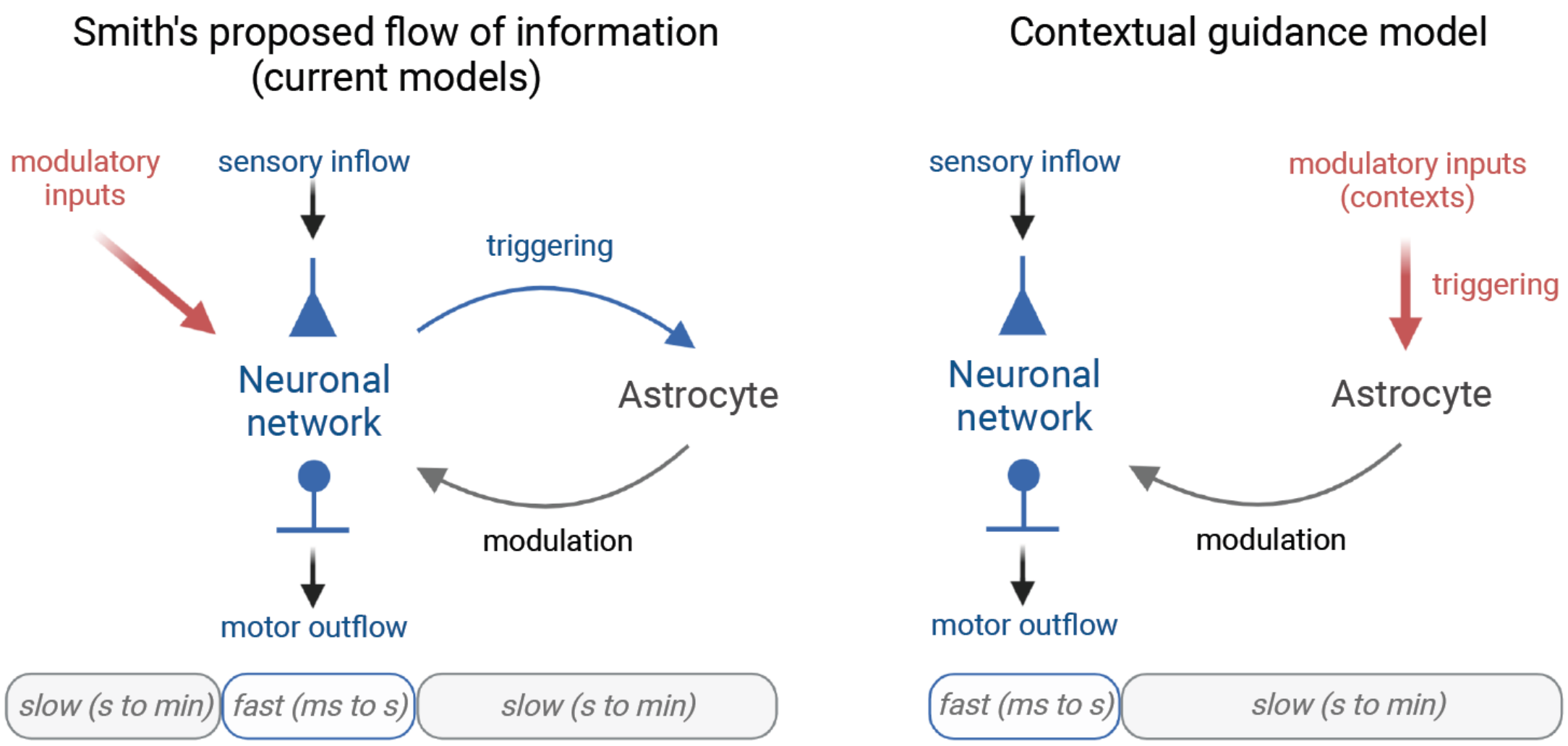

***Figure 4***: **Contextual guidance reutilizes the fundamental constituents of current models.** Contextual guidance is made of the same fundamental parts as current models that theorize astrocyte functions. In current and original concepts (*left*), the contribution of astrocytes to neuronal networks is constrained in a feedback loop: synaptic/neuronal signaling is the main driver of astrocytes' activity which outputs back onto the same (or different) synapses/neurons. Modulatory inputs are assumed to act directly on synapses/neurons. On the contrary, contextual guidance (*right*) opens the source of inputs that engage astrocytes to modulatory signals, which are not synaptic by default. This, in turn, positions astrocytes as the gateway for external inputs into neuronal circuits, *de facto* describing astrocytes as circuit effectors in a feedforward mechanism rather than homeostatic entities. Astrocytes' ability to modulate synaptic properties and firing activity becomes part of the mechanism through which volume-transmitted modulatory signals achieve their effects on neural circuits. This hierarchical architecture reconciles the slowness of astrocytes and modulatory signaling with the fast time scale of neuronal and synaptic activity in a parsimonious fashion (adapted from [5]).

### *f. Context-dependent regulation of metabolism and blood flow*

Another fundamental role of astrocytes, conserved across brain regions and species, involves the regulation of neuron metabolism. This includes both the provision of energetic substrates, most elegantly described by the astrocyte-neuron lactate shuttle hypothesis [4], as well as fatty acid metabolism [66]. Both of these processes function to spare neurons from excessive oxidative stress as astrocytes are endowed with a greater redox capacity than their neuronal counterparts [67]. While neurons possess the machinery necessary for glucose and fatty acid metabolism, they remain dependent on astrocytes for much of these functions [68] and, under conditions of high neuronal activity, astrocytes are required to shoulder some of the metabolic burden and contribute to energy substrate delivery (e.g., L-lactate [69]) and clearance of waste metabolites (e.g., lipid droplets [70]). Interestingly, this metabolic support is shaped by context [55], indicating that energetic and metabolic fluxes through astrocyte networks can, too, be deployed as contextualizers to influence circuit activity. For instance, astrocytic delivery of lactate in the lateral hypothalamus was shown to fluctuate across the 24hr period, influencing orexin/hypocretin neuron activity and directly regulating sleep-wake cycles in mice [69] (**Table 2**). Another study showed that a short inescapable stress episode induced a sustained, glucocorticoid-dependent decrease in astrocyte network function, reducing the capacity of individual astrocytes to meet the energetic demands of nearby neurons, evidenced by decreased synaptic plasticity [24]. This aligns with evidence that corticosterone alters astrocytes coupling [71] and that decoupling astrocytes impairs plasticity and learning [72]. This, in turn, raises the possibility that circadian fluctuations in glucocorticoids affect neural networks by tuning astrocyte metabolic/network function.

Astrocytes are also essential on a broader metabolic scale, such as in regulating feeding behaviors. Like neurons, they express an abundance of receptors for distinct metabolically active molecules [73] including insulin [74], leptin [75], ghrelin [76], or CCK [77] and, upon activation of these receptors, astrocytes mobilize contextualizers to actively regulate neuronal output in various feeding centers of the brain [74,75]. Many of these and other metabolic signals are produced in the periphery and travel to the central nervous system via the vasculature, which is ensheathed by astrocyte processes. As such, astrocytes get exposed to these metabolic signaling molecules in advance of neurons, and are in a position to play pivotal roles in mediating or coordinating central adaptation to changes in metabolic status at large.

Lastly, astrocytes also regulate cerebral blood flow with specialized perivascular endfeet, another one of their defining features [12]. While this is too wide and complex a topic to discuss here, recent work showed that astrocyte's influence over blood vessel dynamics is contextually defined. For instance, Tran et al. [79] found that behavioral states (volitional running and whisking) play a critical role in the extent of endfeet calcium dynamics that sustain arteriole dilation [80], suggesting that astrocytic control of local blood supply is, in essence, an astrocytic contextualizer.

## 4- *New horizons and outstanding questions*

The immediate ramifications of contextual guidance are many. First, it abrogates the idea that astrocytes play "homeostatic functions" as it implies that even the most fundamental function, such as $[K^+]_{ex}$ uptake, is adjusted on-demand to shape the neuropil [30] rather than maintained at a fixed set point. Second, it provides a foothold to explore the determinants of astrocytes-derived transmitter release and understand the manifold nature of astrocytes' responses (i.e., their aptitude to secrete a multitude of signaling molecules). Third, it sheds light on the notions of domains and networks of coupled astrocytes, and their precise anatomical borders. Fourth, it permits a better interrogation of the contribution of astrocytes to network functions, especially as it relates to the reconfiguration of neural circuits produced by neuromodulators [81,82]. Fifth, it provides a functional paradigm that articulates the rapid millisecond time scale of neuronal networks and the second-to-minute time scale of astrocyte activity, reconciling an apparent temporal dilemma (**Fig. 4**). Sixth, it opens better perspectives to probe and understand the role of astrocytes in behavior.

From this theory also ensues a new generation of inquiries. What is the astrocyte "code" linking contextual triggers to contextualizers across circuits, and how does the $Ca^{2+}$-dependency of astrocyte signaling permit input-output specificity/fidelity? Is there an astrocyte memory of circuit configurations (e.g., weight distribution) re-implemented on demand, and can astrocytes get "stuck" in a context? Are astrocytes the gateway to understand the nervous system correlates of behavioral adaptation and cognitive flexibility? To tackle these questions and fully capture the environmental conditions that mobilize astrocytes, the wide adoption of in vivo studies in behaving animals seems inevitable, in line with recent trends.

Lastly, contextual guidance shall be useful to develop new computational theories and address modern artificial intelligence (AI) challenges. The vast majority of efforts to understand and emulate neural computation have indeed focused on circuit mechanisms at the level of neurons and synapses alone, with information represented in spiking patterns. A framework in which astrocytes are conceptualized as orchestrators of information processing, via the slow and systematic modulation of neuronal topology and dynamics, might naturally overthrow these simplistic views. It could, for instance, transform how we theorize about learning and adaptation by animals and humans in dynamic contexts, and over multiple time scales, the computational basis of which remains elusive. In fact, in the field of AI, artificial neural networks (ANNs) and algorithmic systems that only rely on neuronal rules of learning (e.g., Hebbian plasticity) are prone to a phenomenon known as catastrophic forgetting [83,84] because previous memory traces are overwritten as new memories are formed. This prevents agents from accruing knowledge spanning long time-scales. To obviate such fragility, ANNs are often endowed with slow memory elements that allow stable representations of prior experience and contexts over time. Resulting ANNs, such as the long-short-term memory (LSTM) network, have been immensely successful in building generative models of tasks with long time dependencies [85]. However, from a neurobiological perspective, they are difficult to reconcile with any known brain mechanisms. Owing to their ability to recognize contexts and reconfigure neural circuits on-demand, astrocytes are here described as a slow memory element modulating and stabilizing synaptic networks over long epochs, i.e. a natural biological substrate of LSTM (**Fig. 5**). Contextual guidance, therefore, opens the gate to glia-inspired AI.

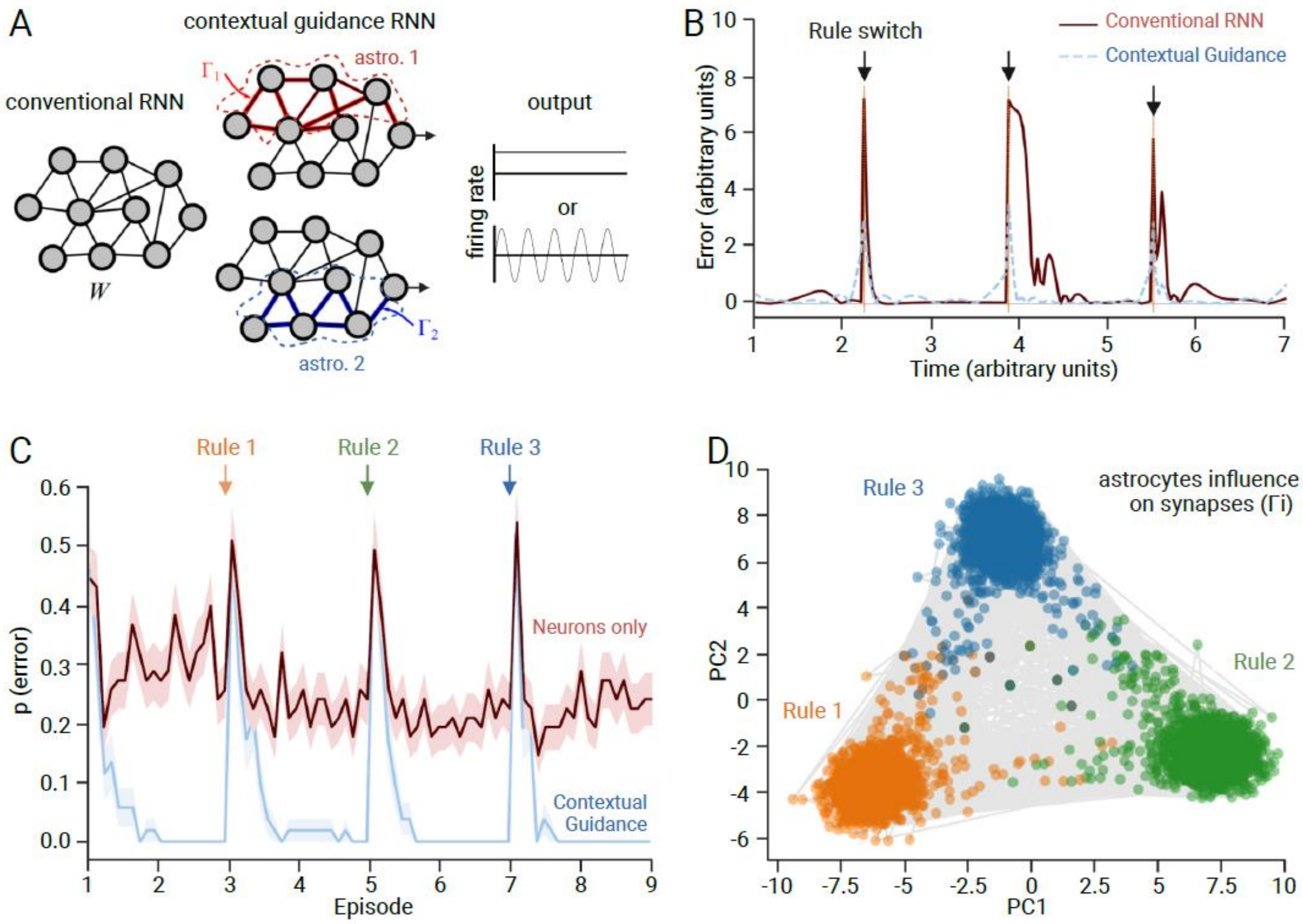

***Figure 5***: **Contextual guidance in recurrent neural networks allows for context adaptation: proof of principle. A**, In a conventional recurrent artificial neural network (RNN), synaptic weights and network connectivity are given by a matrix *W* and modulated by activity, context and prior experience (*left*). In a "Contextual Guidance" RNN (*center*), synaptic weights are given by the sum *W*+*Γi*, where *Γi* reflects the direct contribution of astrocytes over two independent ensembles of connections (red and blue), enacting tiled astrocytic domains (doted areas). The RNNs are tasked with producing a specific output (*right*): bursting activity (harmonic oscillation) or tonic firing, as in [29]. **B**, RNNs are asked to change their output at the occurrence of a contextual signal (arrow), with error measured as a deviation from the expected frequency. In this abstraction, where information about the context (expected output) is known to the RNNs, a conventional RNN has to "re-learn" *W* in each context, resulting in high error transients. The "Contextual Guidance" RNN, however, is able to change its regime rapidly. In effect, a single *W* can be learned such that the network adapts its output by modulating synaptic weights through astrocytes (*Γi*), without resorting to re-learning. **C**, Two LSTM RNNs were built as in A, and trained on an abstraction of the Wisconsin Card Sorting Task (WCST). In the WCST, agents receive a set of cards, each with different numbers, shapes and colors, and are tasked with picking one card to match a hidden rule (e.g. select the card with triangles). In this task, the rule (i.e. context), can only be learned by trial and error, and changes over the course of the task, regardless of the agent's performance. The networks are trained to maximize future reward and learn to determine and exploit a new rule. We used 64 neurons and 16 astrocytes, and numerically optimized all free parameters using episodes of 80 trials. Solid traces show mean error probability and shades shows s.e.m.. The "Contextual Guidance" LSTM-RNN outperforms a conventional network by adapting to the new context and performing optimally until the next context switch. **D**, Dimensionality reduction over the 16 astrocytic outputs (*Γi*) across all trials shows that astrocytic modulation of synaptic weights is highly context-specific, reminiscent of the contextualizing rules described in the main text. Simulations were ran on Matlab (**B**) and Python (**C&D**).

## 5- *Concluding remarks*

Glia have emerged as a topic of great interest to a wider range of neuroscientists, perhaps because they appear more diverse, complex, and active than previously thought. This has most recently been fueled by a series of discoveries placing astrocytes at the center of neuromodulatory pathways, across brain regions and animal models. This provides a resounding call to re-conceptualize the core function of astrocytes, and to understand how 40% of the brain contributes to information processing and behavior in the nervous system besides its electrical passivity. The view presented here depicts astrocytes as multiplexes that decode multiple environmental factors to fine-tune neural circuits in their domain and orchestrate network function in a context-specific, feedforward and hierarchical fashion (**Fig. 4**). Combined with its ability to integrate many features of astrocytes and reconcile disparate observations, contextual guidance, we hope, will prove useful to future inquiries in the fast-expanding field of astrocyte biology, encourage a "glia adoption" by other fields, and guide a multicellular understanding of brain function and behavior.

**Acknowledgments:** C.M.R. was supported by the Canadian Institutes of Health Research Project Grant, NSERC Discovery Grant, Fonds de Recherche du Québec – Santé, and by the Brain & Behavior Research Foundation (NARSAD Young Investigator Award). S.C. was supported by the DoD W911NF-21-1-0312 and the National Science Foundation 1653589. T.P. was supported by the NIH (1R01MH127163-01), the DoD (W911NF-21-1-0312), the Brain & Behavior Research Foundation (NARSAD Young Investigator Award #28616), the Whitehall Foundation (2020-08-35), and McDonnell Center for Cellular and Molecular Neurobiology Award (22-3930-26275U). Authors thank Dr. Jaclyn Dunphy for her critical feedback. We apologize that the work of many of our colleagues and peers could not be cited owing to space limitations. Figures were created with BioRender.com.

**Data Availability Statement**: The datasets generated during and/or analyzed during the current study are available from the corresponding author upon request.

**Code availability statement**: Codes and mathematical algorithms used in Figure 5 will be made available upon publication

**Contribution:** T.P. wrote first drafts, made Fig.1-4, Box 1 and Tables; C.M-R. and S.C. contributed sections along lines of expertise; C.M-R expanded and revised manuscript; S.C. contributed Fig.5; T.P. assembled comments from all authors and wrote the final version.

| ***Responsiveness to*** | ***Body of literature*** | ***Model*** | ***In brief*** |
|---|---|---|---|
| ***Neuromodulators*** *(vigilance and behavioral states)* | Chen, PNAS 2012. [96] | Mouse | Astrocytes sense attention-associated **acetylcholine** to enable stimulus-specific plasticity |
| | Navarrete, PLos Biol 2012. [111] | Mouse | **Acetylcholine**-mediated synaptic plasticity operates via astrocytes in vivo |
| | Pabst, Neuron 2016. [113] | Mouse | Astrocytes sense and mediate the effect of **acetylcholine** neuromodulation in the dentate gyrus |
| | Papouin, Neuron 2017. [26] | Mouse | Astrocytes transform **acetylcholine** neuromodulation into D-serine gating of NMDARs |
| | Takata, J Neuro 2011. [39] | Mouse | Astrocytes transform **acetylcholine** neuromodulation into plasticity |
| | Corkrum et al., Neuron 2020 [97] | Mouse | Astrocytes transform **dopamine** neuromodulation into adenosine tone |
| | Galloway, J Neuro 2018. [103] | Mouse | **Dopamine** remodels astrocyte morphology |
| | Jennings, Glia 2018. [105] | Mouse | **Dopamine** elicits Ca2+ responses in astrocytes |
| | Agarwal, Neuron 2017. [86] | Mouse | **Norepinephrine** elicits Ca2+ signals from astrocyte mitochondria |
| | Bekar, Cereb Cortex 2008. [89] | Mouse | α1NAR triggers cortical astrocyte Ca2+ in vivo |
| | Ding, Cell Calcium 2013. [99] | Mouse | α1NAR mediates Ca2+ signaling in cortical astrocytes |
| | Gordon, Nat Neuro 2005. [104] | Mouse | **Norepinephrine** induces ATP-release from hypothalamic astrocytes |
| | Kohro, Nat Neuro 2020. [37] | Mouse | **Norepinephrine** elicits D-serine release from Hes5+ dorsal horn astrocytes in the spinal cord |
| | Kwak, Neuron 2020. [106] | Mouse | **Norepinephrine** release controls GABA synthesis and release by thalamic astrocytes |
| | Ma, Nature 2016. [107] | Fly larvae | **Norepinephrine** elicits Ca2+ responses in astrocytes which mediates chemotaxic response |
| | Monai, Nat Com 2016. [108] | Mouse | tDCS elicits astrocytes Ca2+ responses via **norepinephrine** & α1NAR |
| | Mu et al. Cell 2019. [109] | Zebrafish | Astrocytes integrate past **norepinephrine** activity to modulate circuit activity and behavior |
| | Oe, Nat Com 2020. [112] | Mouse | **Norepinephrine** is temporally integrated by astrocytes |
| | Pankratov, Frontiers Cell Neuro 2015. [38] | Mouse | **Norepinephrine** induces D-serine and ATP release from neocortex astrocytes via α1NAR |
| | Paukert, Neuron 2016. [34] | Mouse | **Norepinephrine** controls engagement of astrocytes in local circuits |
| | Sherpa, Synapse 2016. [116] (also see [53]) | Mouse | **Norepinephrine** expands astrocytes processes and reduces extracellular space volume |
| | Wang, Science sig 2012. [30] | Mouse | **Norepinephrine** elicits astrocyte-mediated elevation of $[K^+]$ and neuronal excitability |
| | Zuend, Nat Metabo 2020. [27] | Mouse | **Norepinephrine** triggers extracellular release of lactate by cortical astrocytes via β-AR |
| | Woton, J Neurosci Res 2020. [120] | Mouse | **Serotonin** affects [K+] regulation to modulate sensory synaptic inputs and somatosensory adaptation |
| | Nam, Cell Rep 2019. [110] | Mouse | Astrocytic **u-opioid** receptor (MOR) elicits Ca2+ responses and glutamate release |
| ***Hormones & neuro-hormones*** *(vigilance, physiological and metabolic states)* | Ameroso et al., 2022. [28] | Mouse | VMH astrocyte sense food intake-dependent **BDNF** levels to suppress feeding |
| | Crosby, J Neuro 2018. [77] | Mouse | **Cholecystokinin (CCK)** gates astrocyte-derived ATP release in response to glutamate via astrocytic CCK2Rs |
| | Murphy-Royal Nat Com 2020. [24] | Mouse | Stress-associated **corticosteroid** alters plasticity by impairing astrocyte network function and lactate supply |
| | Xia, Neuropharma 2018. [71] | Mouse | **Corticosterone** modifies astrocytes Cx43-connectivity |
| | Fuente-Martín, Sci Rep 2016. [76] | Mouse | **Ghrelin** boosts glutamate transport and reduces glucose uptake by hypothalamic astrocytes via GHSR-1A |
| | Cai, et al. J Clin Invest 2018. [95] | Mouse | **Insulin** triggers ATP release from NAC astrocytes to regulate DA signaling |
| | Garcia-Caceres, Cell 2016. [74] | Mouse | **Insulin** controls hypothalamic POMC activity and connectivity via astrocytes |
| | Kim, Nat Neurosci 2014 [75] | Mouse | **Leptin** signaling in astrocytes influences feeding behavior |
| | Theodosis, Nature 1986. [117] | Mouse | **Oxytocin** induces astrocytes morphological reorganization in SNO |
| | Wahis, Nat Neuro 2021. [22] | Mouse | **Oxytocin** triggers D-serine release from amygdala astrocytes |
| | Park, Curr Biol 2022. [114] | Fly | **Thirst** triggers D-serine release by astrocytes as part of a regulatory loop |
| ***Biochemical context*** *(physiological and metabolic state)* | Angelova, J Neuro 2015. [87] | Mouse | Astrocytes detect **pO2** in the brainstem CPG for breathing |
| | Beltran-Castillo, Nat Com 2017. [91] | Mouse | Elevated **pCO2** elicits a feedback loop through astrocytic D-serine signaling |
| | Gourine, Science 2010. [23] | Mouse | Astrocytes sense **pH/pCO2** in brainstem CPG for breathing |
| | Bouyakdan, JCI 2019. [92] | Mouse | **Feeding**-induced Acyl-CoA Binding Protein (ACBP) release by astrocytes reduces feeding |
| ***Sleep/Wake*** *(vigilance states)* | Bellesi, BMC Biol 2015. [90] (see also [53]) | Mouse | **Vigilance states** remodel astrocytes ultrastructure and transcriptomic profile |
| | Blum, Current Biol 2021. [48] | Fly | Astrocyte calcium encodes **sleep need** |
| | Bojarskaite, Nat Com 2020. [35] | Mouse | Astrocyte activity is **state-dependent** and modulates slow wave sleep |
| | Ding et al, Science 2016. [54] | Mouse | **Sleep/wake** cycling is governed by interstitial space Ionic composition** |
| | Ingiosi, Curr Biol 2020. [49] | Mouse | Astrocyte activity is **state-dependent** and modulates slow wave sleep |
| | Papouin, Neuron 2017. [26] | Mouse | Astrocytes tune NMDAR activity to **wakefulness** |
| | Poskanzer, PNAS 2016. [50] | Mouse | Astrocytes anticipate and orchestrate cortical **states switch** |
| | Schmitt, J Neuro 2012. [115] | Mouse | Astrocytes adenosine release depends on **sleep pressure** |
| | Thrane, PNAS 2012. [118] | Mouse | Astrocytes Ca2+ activity responds to **anesthetics** |
| | Xie, Science 2013. [53] | Mouse | Interstitial space volume and fluid exchange increase during **sleep**** |
| ***Sensory inputs*** *(behavioral states)* | Chen, PNAS 2012. [96] | Mouse | Astrocytes enable **stimulus-specific** plasticity |
| | Cui, Nature 2018. [98] | Rat | **Aversive experience** increases Kir4.1 expression in lateral habenula astrocytes, reducing extracellular [$K^+$] |
| | Doron, Nature 2022. [100] | Mouse | In vivo astrocytes Ca2+ ramps up with proximity to **reward location** |
| | Duan, Neuron 2020. [101] | C. elegans | AMsh glia sense **aversive odorants** and yield olfactory adaptation |
| | Fernandez-Abascal, Neuron 2021. [102] | C. elegans | AMsh glia mediates **touch response** |
| | Morquette, Nat Neuro 2015. [29] | Mouse | Astrocytes respond to **orofacial sensory inputs** to induce mastication |
| | Paukert, Neuron 2015 [34] | Mouse | V1 astrocytes do not respond to **visual stim** except during locomotion |
| | Ding, Cell Cal 2013. [99] | Mouse | V1 astrocytes do not respond to **visual stim** except during locomotion |
| | Takata, J Neuro 2011. [39] | Mouse | Astrocytes enable **stimulus-specific** plasticity |
| | Tran, Neuron 2018. [79] | Rat | Astrocytes Ca2+ activity integrates **behavioral states** and vascular signals |
| ***Circadian time*** *(behavioral and vigilance states)* | Barca-Mayo, Nat Com 2017. [88] | Mouse | **Bmal1 activity** in astrocytes drives circadian locomotor activity and cognition |
| | Brancaccio, Neuron 2017. [94] | Mouse | Astrocytes control **circadian timekeeping** |
| | Brancaccio, Science 2019. [83] | Mouse | **Cell-autonomous clock** in astrocytes drives circadian behavior |
| | Tso, Curr Biol 2017. [119] | Mouse | Astrocytic **Bmal1** protein controls daily rhythms in suprachiasmatic nucleus |

***Table 1***: **Body of work documenting the wide responsiveness of astrocytes to a variety of molecules and contexts**. Kir4.1: inward-rectifying potassium channel, tDCS: transcranial direct-current stimulation, α1NAR: alpa-1 norepinephrine receptor. **denotes studies where the role of astrocytes is highly suspected, but was not directly demonstrated.

| *Reference* | *Brain area* | *Function* | *Context* | *Contextual trigger* | *Astrocyte sensor* | *Contextualizer and Circuit tuning* | *Contextual adaptation* |
|---|---|---|---|---|---|---|---|
| *Ameroso et al. 2022 [28]* | Mouse ventromedial hypothalamus (VMH) | Regulation of energy balance | Positive caloric status (nutrient availability) | BDNF | TrkB.T1 receptor | Decreased PAP coverage of excitatory synapses → reduced glutamate reuptake → increased VMH neuron activity | suppression of feeding behavior and increased glycemic control → maintenance of energy homeostasis |
| *Angelova et al.2015 [87]* | Mouse and rat pre-Bötzinger complex of the ventral medulla oblongata | Brainstem CPG for breathing | Hypoxia | Decrease in partial pressure of O2 (pO2) | Mitochondrial respiration | Increased ROS production → increased lipid peroxidation → PLC activation → IPR2 activation → Ca2+ activity → vesicular release of ATP | Increased breathing frequency → normalization of pO2 |
| *Beltran-Castillo et al., 2017 [91]* | Mouse caudal medullary brainstem | Central chemoreception | Hypercapnia | Increase in partial pressure of CO2 (pCO2) | Unknown pCO2 sensor | D-serine release in raphe nucleus and ventral respiratory column → activation of NMDARs | Increased breathing frequency → normalization of pCO2 |
| *Brancaccio et al. 2017 & 2019 & (see also Tso, et al. 2017) [93, 94,119]* | Mouse suprachiasmatic nucleus (SCN) | Central circadian time-keeping and pacemaker | Circadian night | unknown | Astrocyte cell-autonomous transcription-translation feedback loop | Cx43-dependent glutamate release → presynaptic GluN2C-NMDARs activation on SCN neurons → increased GABAergic tone → synchronization of dorsal SCN neurons → drives clock gene expression in SCN neurons | Instructs SCN circadian clocks and orchestrates circadian behavior |
| *Chen et al., 2012 [96]* | Mouse visual cortex (V1) | Processing of primary visual inputs from the thalamus | Stimulation of nucleus basalis paired with visual input (orientation gratings) | Acetylcholine (ACh) neuromodulation | Muscarinic ACh receptor (AChR) | Ca2+ activity → NMDAR-dependent depolarization of V1 excitatory neurons → selective potentiation of visual responses specific to the orientation used for paired visual stimulus | Stimulus (orientation)-specific potentiation → improved orientation-specific responses in V1 |
| *Cui et al. 2018 [98]* | Rat lateral habenula | Learning of aversive experience | Unescapable aversive challenge and congenital learned helplessness | Unknown | Unknown | Increased Kir4.1 expression → decreased [K+]ex → neuronal hyperpolarization → de-inactivation of L-type voltage-sensitive Ca2+ channels → NMDAR-dependent bursting → suppression of monoaminergic centers | Learned helplessness (“give up” behavior as in [109]?) |
| *Duan et al., 2020 [101]* | C. elegans peripheral sensory system | Olfactory receptive organ | Aversive environment | isoamyl alcohol (IAA) and other aversive odorants | Amphid sheath (AMsh) glial cell odorant GPCR (SRH-79) | Ca2+-dependent GABA release → hyperpolarization of ASN sensory neuron → inhibition of ASN and reduced responses to aversive odorant | Suppression of aversive odorant-triggered avoidance → olfactory adaptation |
| *Garcia-Caceres et al. 2016 [74]* | Mouse dorsomedial hypothalamus | Glucose sensing & systemic glucose homeostasis | Hyperglycemia | Elevated insulin levels | Insulin receptor (IR) | Glucose entry in CNS → glucose-induced activation of hypothalamic POMC neurons + changes in circuity connectivity → glucose-induced circuit response | Increased CNS and systemic glucose metabolism |
| *Gourine et al. 2010 [23]* | Rat medulla oblongata | Brainstem respiratory chemoreceptor | Hypercapnia | Drop in local pH (due to elevated pCO2) | Unknown H+ sensor | ATP release → activation of P2Y1R and depolarization of retrotrapezoid chemoreceptor neurons → increased phrenic nerve activity | Increased breathing frequency → normalization of pCO2/pH |
| *Kwak et al., 2020 [106]* | Mouse thalamus ventrobasal nucleus | Tactile sensory gate keeper | Attentive state | Norepinephrine (NE) neuromodulation | α1-AR | α1-AR → PLC pathway activation → Ca2+ elevation → Ca2+-dependent activation of Best1 → enhanced GABA release → post-synaptic shunting of lemniscal inputs → inhibition of thalamo-cortical (TC) neuron firing probability | Increased dynamic range of TC neurons stimulus-response + increased temporal fidelity of TC neurons → augmented sensory acuity in attentive states |
| *Ma et al. 2016 [107]* | Fly larvae | N/A | Olfactory or mechanosensory stimulation | Tyramine /Octopamine neuromodulation | Oct-TyrR | ATP release (suspected) → activation of presynaptic A1R on dopaminergic neurons → dopamine neurons silencing | Olfactory-driven chemotaxis and touch-induced startle response |
| *Morquette et al. 2015 [29]* | Rat trigeminal main sensory nucleus (NVsnpr) | CPG for mastication | Palatable food in mouth (inferred) | Stimulation of orofacial sensory inputs | Unknown | S100B release → $[Ca^{2+}]_{ex}$ buffering → de-inactivation of $I_{NaP}$ conductance → rhythmic burst firing of NVsnpr neurons | Mastication rhythmogenesis motor command for mastication |
| *Mu et al. 2019 [109]* | Zebrafish lateral medulla oblongata (MO) | Control of motor behavior | Repeated failed swim attempts (visuomotor mismatch ) | Norepinephrine neuromodulation | α1B-AR | Increased Ca2+ activity in "radial" astrocytes → activation of GABAergic edge cells in lateral MO→ inhibition of motor command | Suppression of swim → futility-induced passivity |
| *Murphy-Royal, et al. 2020 [24]* | Mouse neocortex and hippocampal CA1 | Learning and memory | Stress | Elevated glucocorticoid levels | Glucocorticoid Receptor | Decreased gap junction coupling → reduction of lactate shuttling from astrocytes to neurons → synaptic energy deficit | Suppression of long-term potentiation of synaptic transmission |
| *Papouin et al. 2017 [26]* | Mouse hippocampal CA1 | Spatial learning | Wakefulness/ high vigilance | Increased acetylcholine tone | α7nAChR | D-serine release → increased activation of NMDARs at CA3-CA1 synapses | Augments hippocampal plasticity and learning during wake |
| *Park et al., 2022 [114]* | Drosophila brain | N/A | Water deficiency | Suspected increased sensitivity to glutamate | astray-encoded phosphoserine phosphatase | D-serine synthesis upregulated → D-serine vesicular release → NMDAR activation → activation of quiescent glutamatergic circuit | Increased water consumption and improved water-seeking behavior |
| *Paukert el al., 2014 [34]* | Mouse V1 | Thalamic visual inputs processing | Locomotion | Locomotion-induced norepinephrine-release | α-AR | NE shifts the gain of astrocyte network | State-dependent contribution of astrocytes to sensory-processing |
| *Takata et al. 2011 [39]* | Mouse barrel cortex | Circuit for primary somatosensory inputs | Attention-driven cholinergic inputs from nucleus basalis paired with sensory inputs | Acetylcholine neuromodulation | mAChR | Ca2+-dependent D-serine release → increased activation of NMDARs | Attention-dependent input-specific synaptic potentiation in somatosensory cortex |
| *Wahis et al. 2021 [22]* | Rat and mouse central amygdala (CeM and CeL) | Control of anxiety and emotional behavior | Negative emotionality | Oxytocin (OT) release from PVN | OT receptor (OTR) (subpopulation) | Ca2+-dependent D-serine release → increased NMDAR activity of CeL interneurons → inhibition of CeM output projection neurons | Reduces anxiety and promotes "positive emotional state" |
| *Zuend et al. 2020 [27]* | Mouse somatosensory cortex | Circuit for primary somatosensory inputs | High arousal | Norepinephrine neuromodulation | NE β-receptors | Lactate release in the extracellular space + mobilization of lactate from glycogen stores | Supply of the energetic substrate lactate to neurons during high arousal states |

***Table 2*: Published examples of astrocytic contextual guidance of neuronal circuits:** twenty studies, or combination of studies, describing molecular pathways and behavioral outcomes that amount to the contextual guidance of the local network structure and/or function by astrocytes.

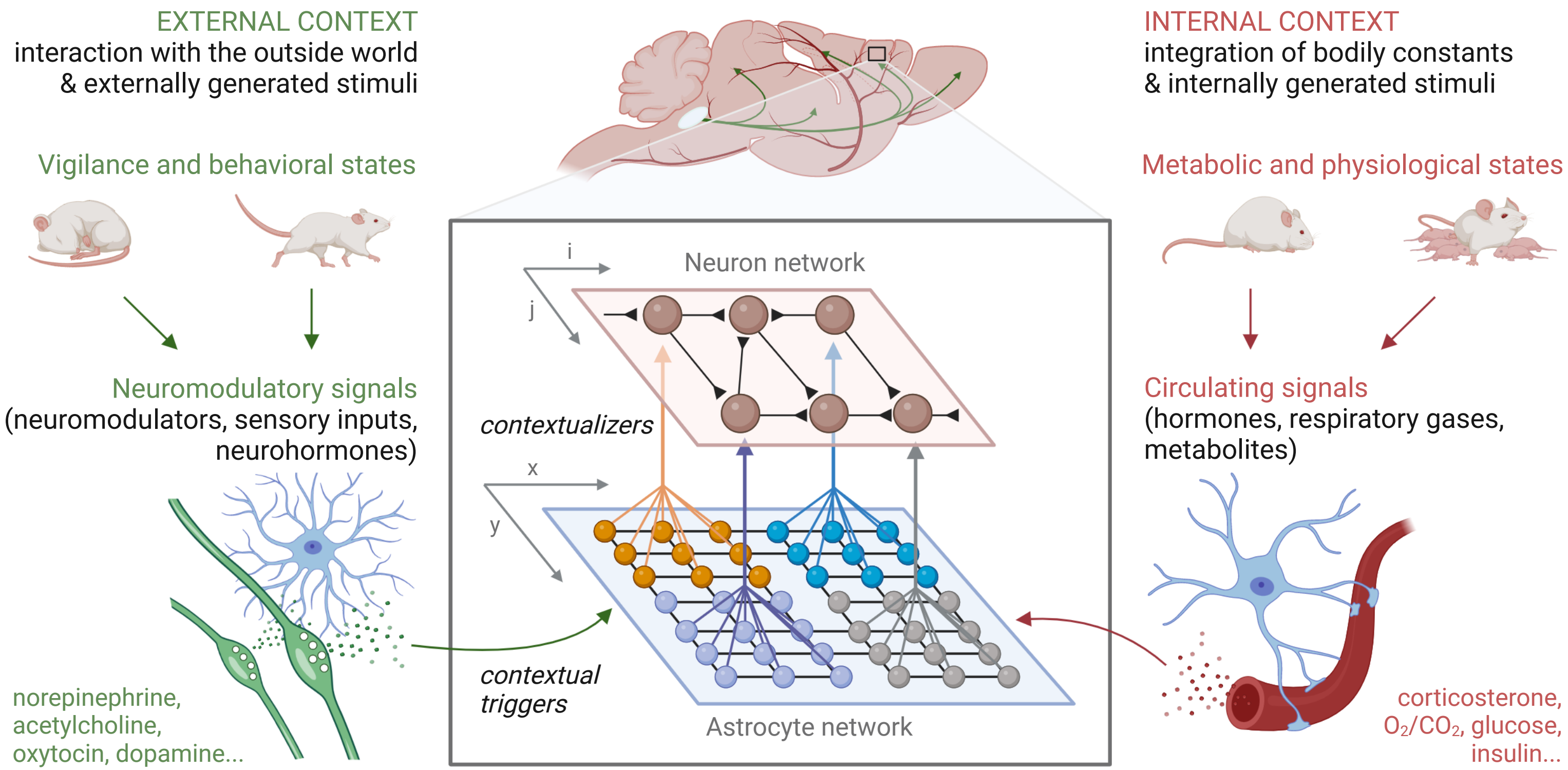
EXTERNAL CONTEXT
interaction with the outside world
& externally generated stimuli
Vigilance and behavioral states
Neuromodulatory signals
(neuromodulators, sensory inputs,
neurohormones)
norepinephrine,
acetylcholine,
oxytocin, dopamine...
INTERNAL CONTEXT
integration of bodily constants
& internally generated stimuli
Metabolic and physiological states
Circulating signals
(hormones, respiratory gases,
metabolites)
corticosterone,
$O_2/CO_2$, glucose,
insulin...
Neuron network
i
j
contextualizers
x
y
contextual
triggers
Astrocyte network

Figure 1

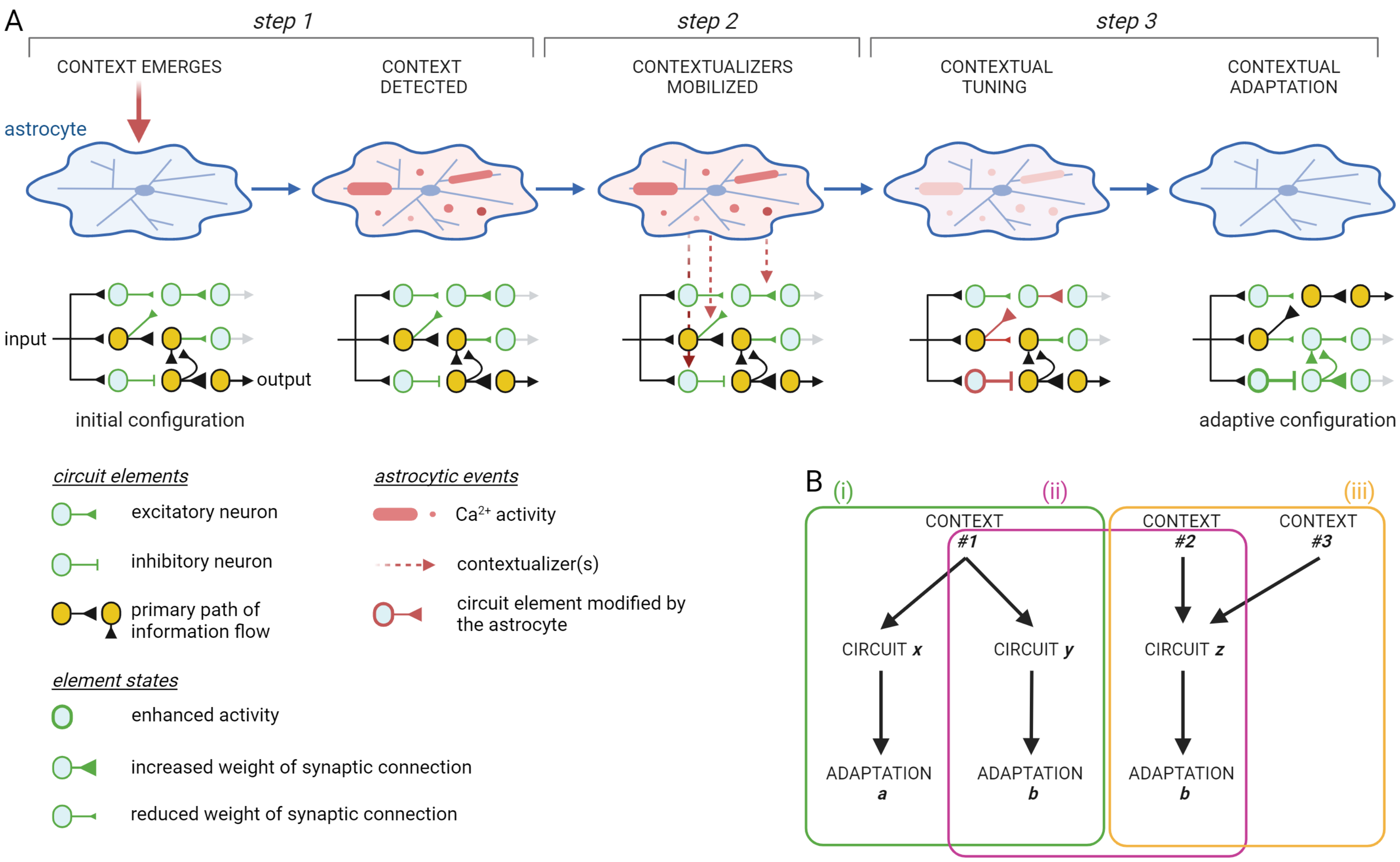

Figure 2

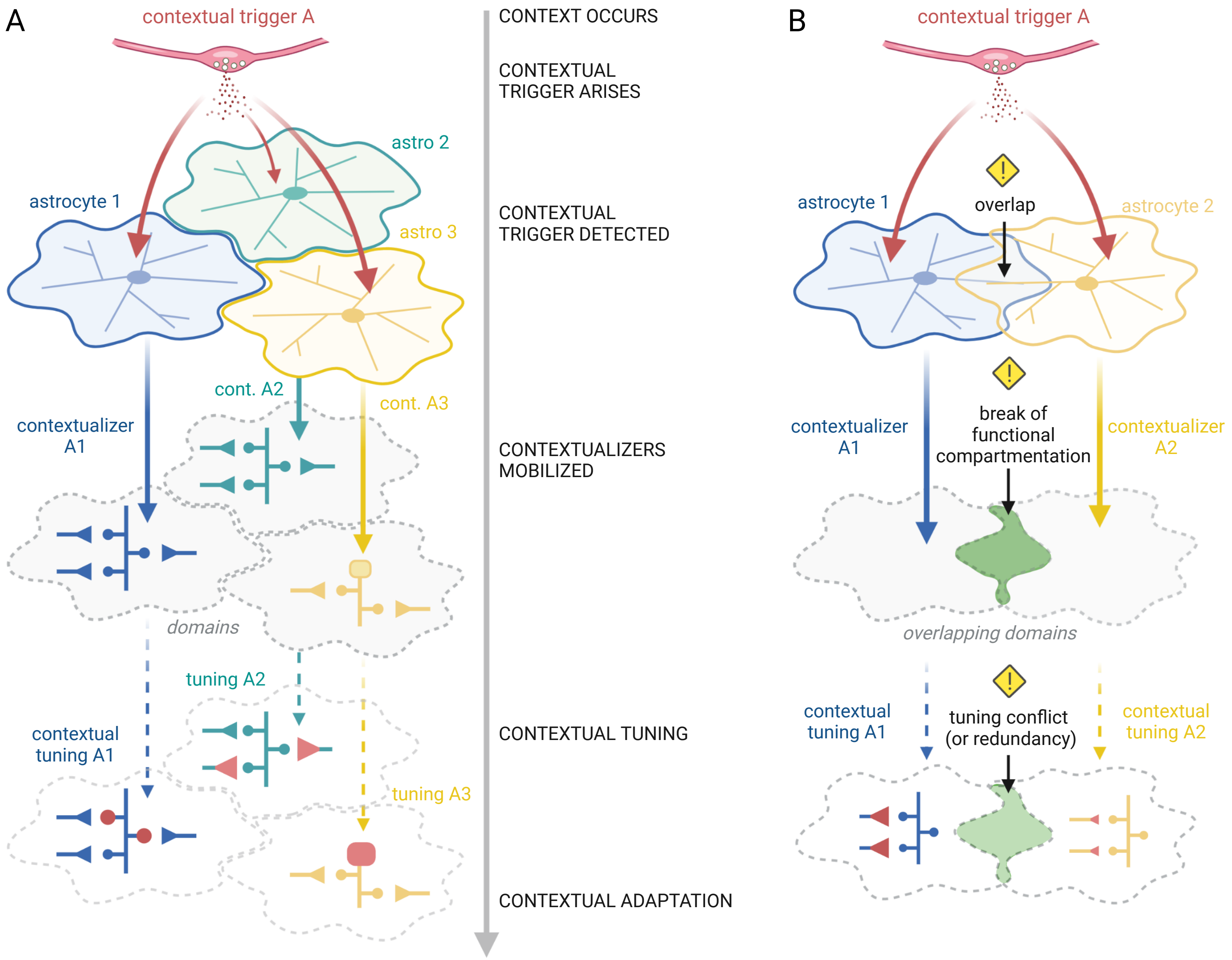

Figure 3

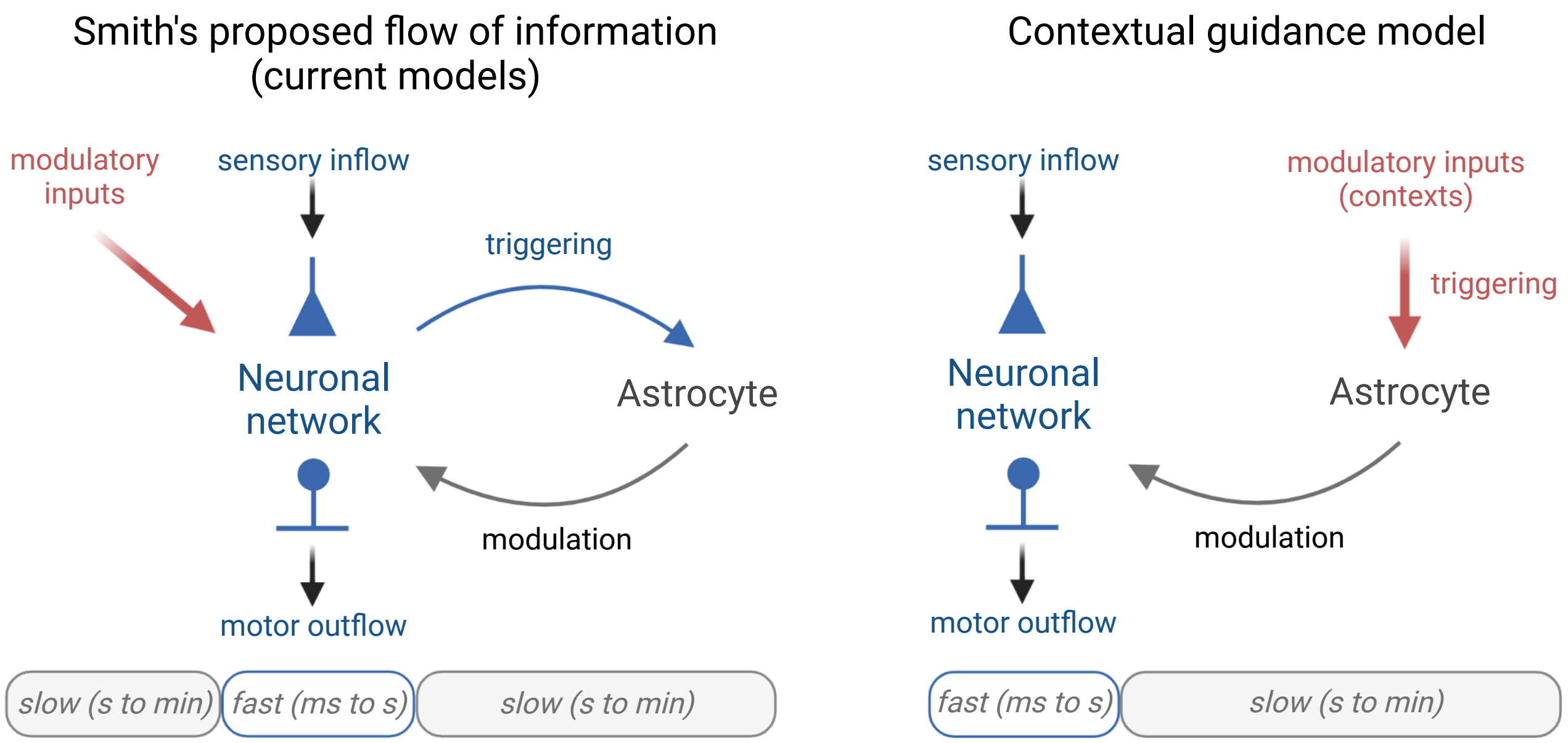

Smith's proposed flow of information
(current models)
modulatory
inputs
sensory inflow
triggering
Neuronal
network
Astrocyte
modulation
motor outflow
slow (s to min)
fast (ms to s)
slow (s to min)
Contextual guidance model
sensory inflow
modulatory inputs
(contexts)
triggering
Neuronal
network
Astrocyte
modulation
motor outflow
fast (ms to s)
slow (s to min)

Figure 4

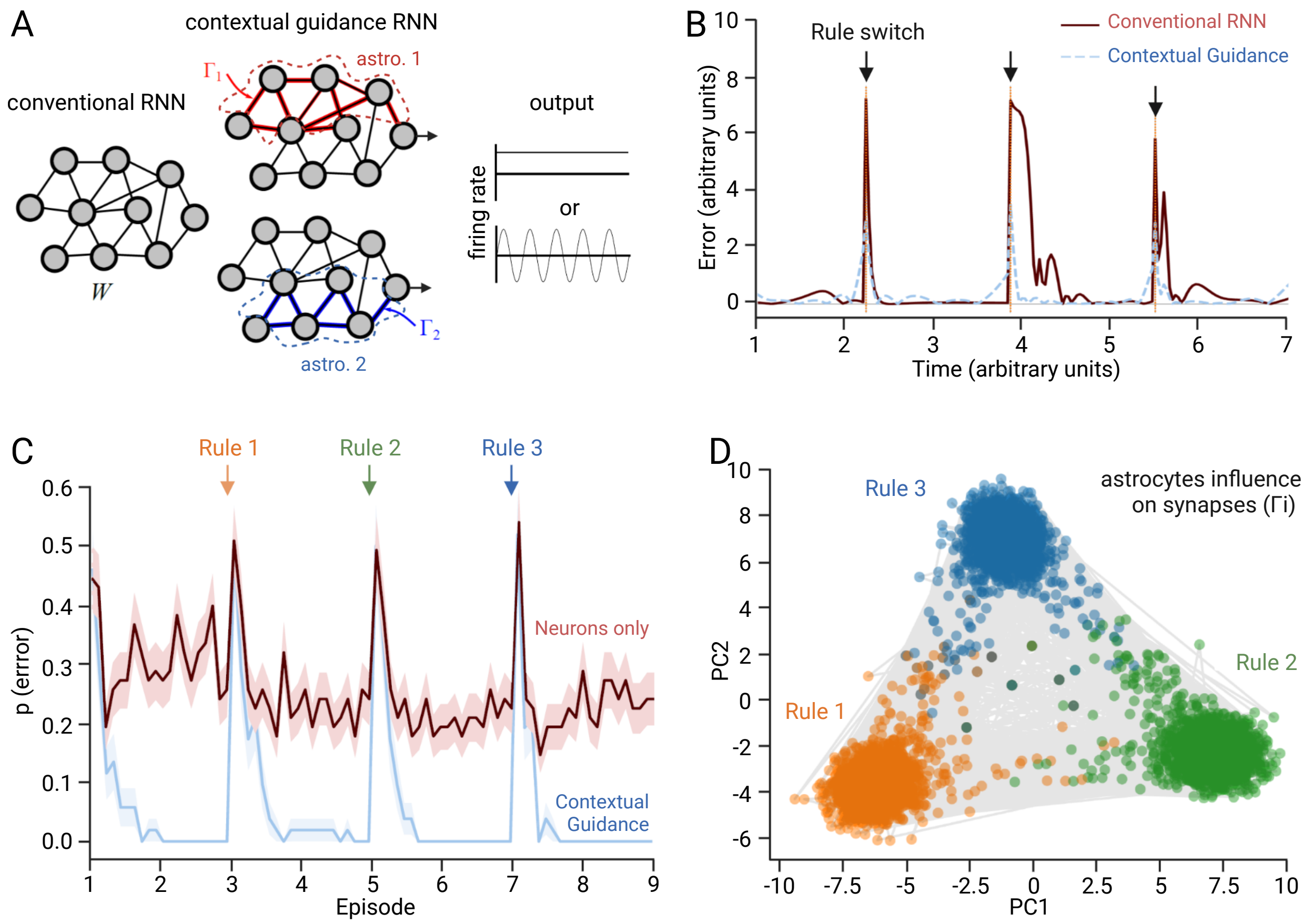

Figure 5

### **Box 1** | **Definitions**

**Context**: ensemble of conditions that relate to an internal or external organismal status, such as a physiological, metabolic, biochemical, behavioral, or vigilance state

**Contextual trigger**: a signal that i) diffuses through the interstitial space, ii) bears significance to the function of the local circuit, iii) encodes a change in internal or external state and, iv) has the ability to mobilize astrocyte signaling

**Contextualizer**: astrocytic output, mobilized in response to a contextual trigger, that is consequential to the underlying neural circuitry

**Contextual tuning**: effect of a contextualizer on the activity and/or functional topology of a neural circuit

**Contextual adaptation**: adaptative value of the circuit reconfiguration produced under contextual tuning, with regard to the context that triggered it

**Contextualizing rules**: association of specific instances of the elemental parts described above into a unique "context→contextual trigger→contextualizer→contextual tuning and adaptation” sequence